\documentclass[twocolumn,showpacs,showkeys,preprintnumbers,amsmath,amssymb]{revtex4}
\usepackage{amsmath}

\usepackage{graphicx}
\usepackage{dcolumn}
\usepackage{bm}
\UseRawInputEncoding

\begin{document}

\def\vsigma{{\hbox{\boldmath $\sigma$}}}
\def\vlambda{{\hbox{\boldmath $\lambda$}}}

\title{ Precision Imaging of the Pion Emission Source in Heavy-Ion Collisions via a Global Rest Frame Analysis \footnote{This work is supported in part by the National Key R\&D Program of China (No. 2024YFA1610700)}}

\author{Qi-Chun Feng$^{1}$}
\author{Yi-Bo Hao$^{1}$}
\author{Yue-Kai Zhou$^{1}$}
\author{Xu Sun$^{2}$}
\author{Yue Jiang$^{1}$}
\author{Jing-Bo Zhang$^{1}$}
\author{Lei Huo$^{1}$}
\author{Yan-Yu Ren$^{1}$\footnote{ryy@hit.edu.cn}}

\affiliation{
$^1$Department of Physics, Harbin Institute of Technology,
Harbin, Heilongjiang 150006, China\\
$^2$ Institute of Modern Physics, Chinese Academy of Sciences, Lanzhou 730000, China\\
$^2$     School of Nuclear Science and Technology, University of Chinese Academy of Sciences, Beijing 100049, China\\
$^2$    State Key Laboratory of Heavy Ion Science and Technology, Institute of Modern Physics, Chinese Academy of Sciences, Lanzhou 730000, China
}

\begin{abstract}
Conventional imaging of pion emission sources, conducted in the center-of-mass frame of individual pion pairs (CMFP), suffers from frame-dependent kinematic distortions that bias the reconstructed source morphology. This method introduces spurious correlations due to the relative boost between the CMFP and the true source rest frame (CMFS), leading to systematic image distortions with a pronounced non-Gaussian tail. We present a transformative approach by performing correlation analysis directly in the global source rest frame (CMFS), the physically meaningful reference frame of the collision fireball. This paradigm shift eliminates kinematic contamination inherent in conventional CMFP-based imaging. The resulting source image shows a dramatic suppression of the non-Gaussian artifact and achieves significantly better agreement with pristine model source functions. Our technique offers a more direct and uncontaminated probe of the intrinsic source geometry, overcoming the limitations of prior methods. It provides a clearer and more accurate determination of the spacetime properties of the nuclear collision fireball, marking a significant advancement in the field.

\end{abstract}



\maketitle

\section{Introduction}

Ultrarelativistic heavy-ion collisions generate nuclear matter with extraordinary energy densities\cite{PRL134011903,NPA1047122874,CPC49-094001,CPC49-043103}. The subsequent decay processes of this matter are profoundly governed by the nuclear equation of state (EOS), potentially involving phase transitions to deconfined states such as quark-gluon plasma (QGP)\cite{EPJC84247,CPC49-094106}. Systems experiencing first-order phase transitions demonstrate substantially greater spacetime expansion compared to those remaining in the hadronic phase\cite{PRL131171404,CPC48-053105,EPJA6115}. Hydrodynamic models verify this expansion pattern for particle emission sources, except in cases where hadronization proceeds through supercooled states\cite{PRL864783}. Furthermore, the spatial configuration of particle emission sources serves as a diagnostic tool for identifying phase transitions and locating potential critical endpoints in a QCD phase diagram\cite{PRC73024903,CPC48-083102}.

Particle interferometric measurements yield crucial insights into the emission source characteristics of particles generated in relativistic heavy-ion collisions\cite{EPJA60135,CPC45-024106,JPG52025102}.
Identical pion pair correlation measurements via HBT interferometry offer unique capabilities for investigating the spatiotemporal characteristics of particle emission sources generated in relativistic nucleus-nucleus collisions\cite{UAW99,RMW00,MAL05,CMS18}. The analysis methods can be summarized into two
categories: Gaussian fitting analysis and imaging techniques. The Gaussian fitting analysis is a model-dependent method, in which one assumes a Gaussian-shaped emission
function prior to extracting the fitting of the HBT parameters
\cite{Cha95,Spr98,UAW99,UHE99}. When the spatial distributions of the particle emission sources generated in ultrarelativistic heavy-ion collisions deviate significantly from the Gaussian profiles
\cite{PHE07,PCH05,PCH07,PCH08,PHE08,RAL08,ZTY09,ZWL02,ZWL04,TCS04,WNZ06,YYR08,PHE18},
this Gaussian fitting analysis is inappropriate\cite{UHE99,TCS04,SNI98,DHA00,EFR06,BAL23}.

In comparison, the imaging methodology developed by Brown and Danielewicz constitutes a model-independent approach\cite{DAB97,DAB98,DAB01}. This technique enables direct reconstruction of the two-pion source function $S(r)$ from the Hanbury-Brown-Twiss correlation data, where $r$ denote the spatial separation between particle pairs. As an established analytical tool, this imaging method has been comprehensively refined for the geometric characterization of emission sources in ultrarelativistic heavy-ion collisions
\cite{PHE07,DAB01,SYP01,GVE02,PCH03,PDA04,PDA05,DAB05,PCH05,PDA07,PCH07,PCH08,PHE08,RAL08,ZTY09,ZWN09}.

Physically, the single particle distribution $S(r)$ is a more direct characterization of the particle-emitting source geometry, in which $r$ represents the distance from a single particle to the source center.
According to probability theory, if $S(r)$ follows a Gaussian distribution, then $S(r)$ must also follow a Gaussian distribution. Analysis of Au+Au collisions at ${\sqrt {s_{NN}}}=200$ GeV reveals that the  extracted pion-emitting source function $S(r)$ exhibits a more prominent tail from the Gaussian form\cite{PHE07}. Experimental findings demonstrate that pion emission originates from both a dense central core and an extended halo region associated with the decays of long-lived resonances\cite{UHE99,PCH05,PCH07,PCH08,RAL08,PHE18}.

This study re-examines the underlying mechanisms responsible for the extended tail observed in pion-emission source distributions. The imaging methodology employed for the analysis intrinsically contributes to tail formation through its fundamental operational principles. Previous investigations utilizing the center-of-mass frame of pion pairs (CMFP) for imaging analysis revealed inherent limitations: varying velocities across different pion pairs introduced geometric distortions and obscured the physical interpretation of reconstructed source functions\cite{PHE07}. To address these systematic effects, we develop an enhanced imaging approach specifically designed for implementation in the source's center-of-mass reference frame (CMFS), which provides more physically meaningful spatial reconstructions.

\section{Traditional analysis of a gaussian source in the CMFP frame}

To establish the reliability of the imaging methodology, a comparative analysis was conducted between the reconstructed source distribution obtained through imaging and the theoretical source distribution derived from the model calculations. \cite{DAB01,DAB05}. The imaging source  $S(r)$ is calculated using an imaging technique. The theoretical source model was constructed by simulating pion pairs emitted from the source, with their relative momentum constrained below $60$ MeV/c, following established methodologies \cite{DAB05,PAN99}.
Now we simulate an emission source with a single-particle distribution in a Gaussian distribution:
\begin{eqnarray}
\label{all bk555}
S(r,t)=
\frac{1}{\pi^2r^3\tau}\exp\left(-{r^2}/{2{r_0}^2}-{t^2}/{2\tau^2}\right).
\end{eqnarray}

The spatial coordinates of the particles were initialized following a Gaussian distribution profile with a characteristic radius ${r_0}=5$ fm femtometers. The emission time duration is parameterized with a Gaussian distribution featuring a temporal width of $\tau=5$ fm/$c$. The pion momenta were assigned according to a Boltzmann distribution with freeze-out temperature ($T_f=130$ MeV).

\begin{figure} [h]
\vspace{10mm}
\includegraphics[angle=0,scale=0.25]{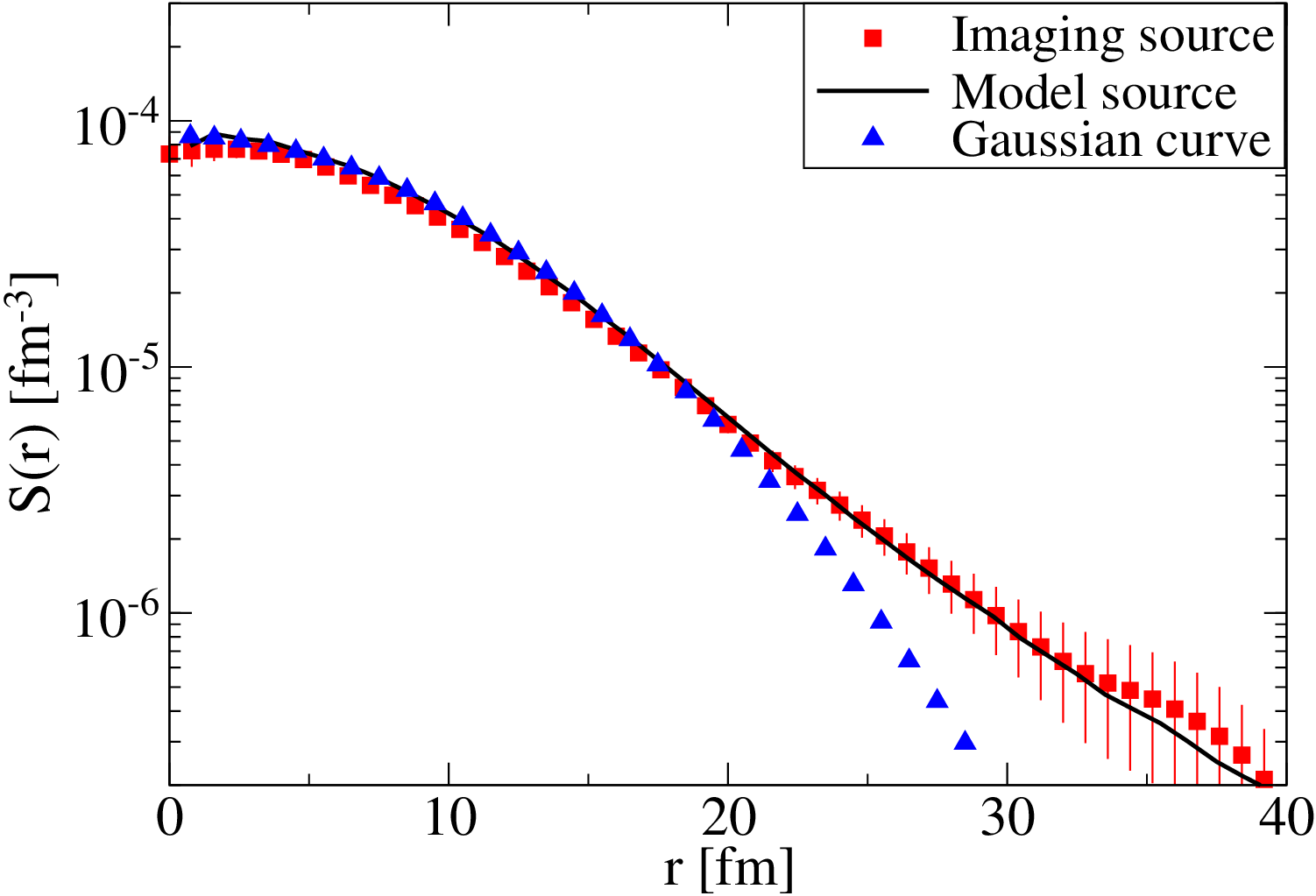}
\caption{(color online). The comparison between the model source(solid curve) and the imaging source($\blacksquare$) with a Gaussian model in the CMFP. A Gaussian curve($\blacktriangle$) is added to check the result.} \label{1}
\end{figure}

The characteristics of the source function in the CMF (Center-of-Mass Frame) framework are illustrated in  Fig.
\ref{1}, which compares the model source with the imaging source using a Gaussian profile. The solid curve representing the model source is derived from all pion pairs generated within the model framework, with their interparticle distances computed in the CMF. The close agreement between the imaging source (squares) and the model source (solid curve) in  Fig.\ref{1} demonstrates the effectiveness of the imaging methodology in the CMF context. Furthermore, based on probabilistic considerations, the source function $S(r)$ is expected to exhibit Gaussian behavior\cite{PHE07,PCH05}.

\section{Reason for the appearance of the tail}


The imaging methodology can be formally generalized as demonstrated in Refs\cite{DAB97,DAB01}. Under the assumption of negligible pion-pion interactions, the two-pion correlation function may be mathematically expressed by  Eq(2)\cite{UAW99}.
\begin{eqnarray}
\label{all bk0} C({\bf q})=1+\int d^4{x}\cos({\bf q}\cdot{\bf
r}-{\bigtriangleup E}{\bigtriangleup t})S(x),
\end{eqnarray}

where, $S(x)$  represents the relative distance distribution. When transformed into the CMFP framework, the term ${\bigtriangleup E}{\bigtriangleup t}$  vanishes, allowing Eq. (\ref{all bk0}) can be rewritten as follows:

\begin{eqnarray}
\label{all bk01} C({\bf q})=1+\int \cos({\bf q}{\cdot}{\bf r})
S({\bf r}) d {\bf r}.
\end{eqnarray}

where the source function$S({\bf r})=\int d(x) dt$ dt characterizes the spatial distribution of the emission point separations for the particle pairs. The angle-averaged form of Eq. (\ref{all bk01}) becomes:
\begin{eqnarray}
\label{all bk1}\mathcal {R}({q})&=&C({q})-1
=4\pi\int\frac{1}{{q}}{\sin({q}r)}{r}S(r)dr.
\end{eqnarray}

The one-dimensional source function may then be reconstructed via Fourier transform:
\begin{eqnarray}
\label{all bk2} S(r)=\frac{1}{2\pi^2}\frac{1}{r}\int\mathcal
{R}({q}){q}\sin({q}{r})d{q}.
\end{eqnarray}
This derivation demonstrates that the imaging technique provides a model-independent method for determining the source function of a CMFP.
However, it should be noted that the CMFP is not a universal reference frame because different pion pairs possess distinct center-of-mass velocities.
Consequently, the source function derived from the CMFP cannot directly characterize the spatial scale of the emission sources.

The above analysis demonstrates that the imaging technique used in the CMFP plays an important role in creating a long tail. To eliminate this effect, technical improvements have been proposed for extracting the source function $S(r)$ from a unified frame.

\section{Extracting two-pion source function in CMFS}

Reference \cite{REN17} introduced an enhanced imaging methodology for source function extraction within a center-of-mass frame system (CMFS) by implementing Equations Eq.(\ref{all bk0})and(\ref{all bk2}) in the reference frame.While Eq. (\ref{all bk0}) remains frame-independent, the non-zero value of $\triangle E \triangle t$ in CMFS introduces inaccuracies in Eq.(\ref{all bk01}). To address this limitation, the improved technique selectively employs pion pairs with minimal energy differences ($\triangle E$) for correlation function construction in CMFS.
This approach effectively mitigates the $\triangle E \triangle t$ perturbation, thereby yielding imaging results that closely approximate the true source function.

\begin{figure} [h]
\vspace{8mm}
\includegraphics[angle=0,scale=0.33]{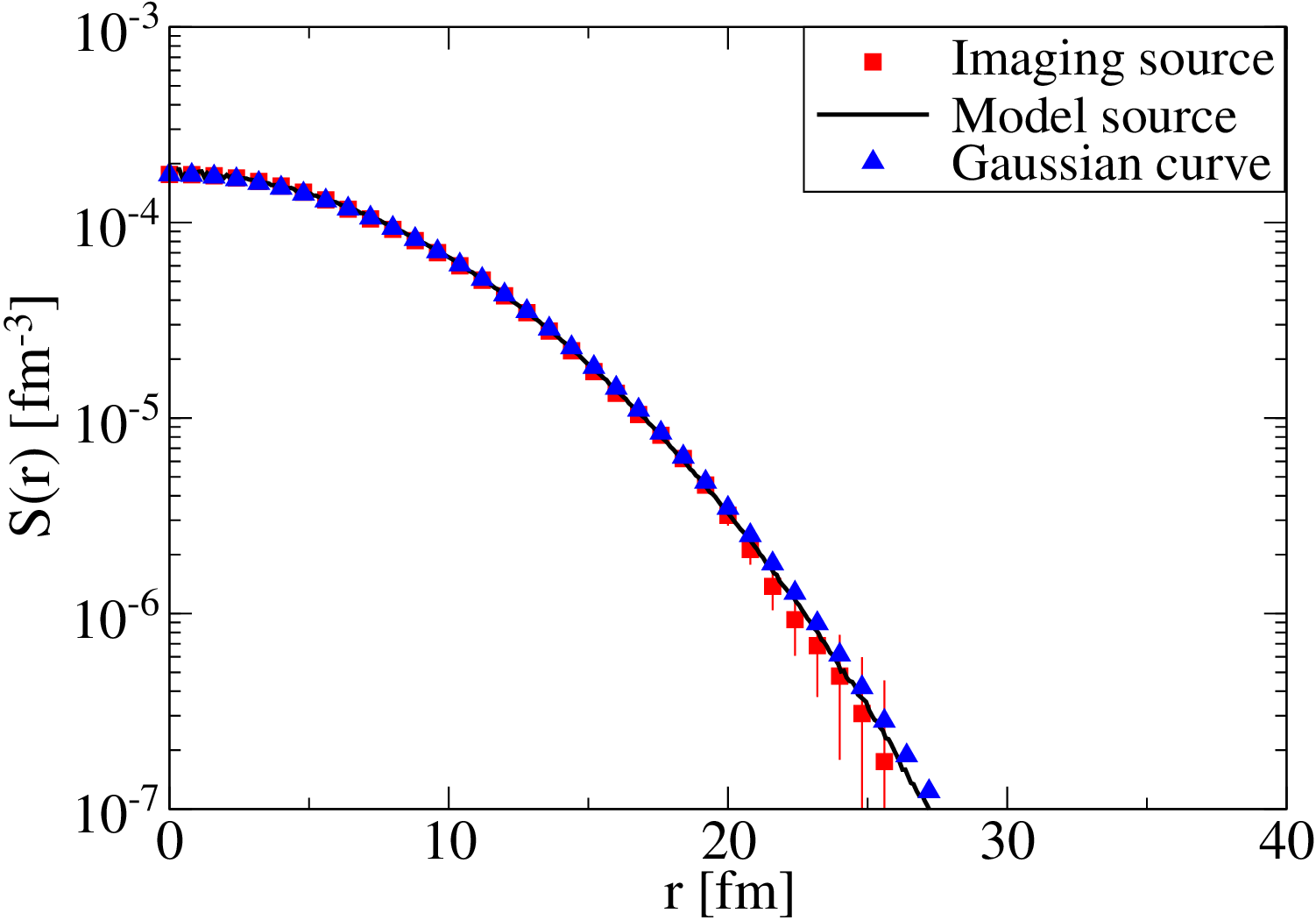}
\caption{(color online). The comparison between the model source(solid curve) and the imaging source($\blacksquare$) with a Gaussian model in the CMFS. A Gaussian curve($\blacktriangle$) is added to check the result.} \label{2}
\end{figure}

For a deeper analysis of the source function characteristics in the CMFS, Fig.\ref{2} presents a direct comparison with $\triangle E$ using CMFS-derived data. The imaging results (square markers) were obtained from pion pairs with energy differences $\triangle E < 10$ MeV, while the model source (solid curve) represents all generated pion pairs with distances calculated in CMFS. The excellent agreement between the curves further validates the methodology.

\begin{figure} [h]
\vspace{12mm}
\includegraphics[angle=0,scale=0.4]{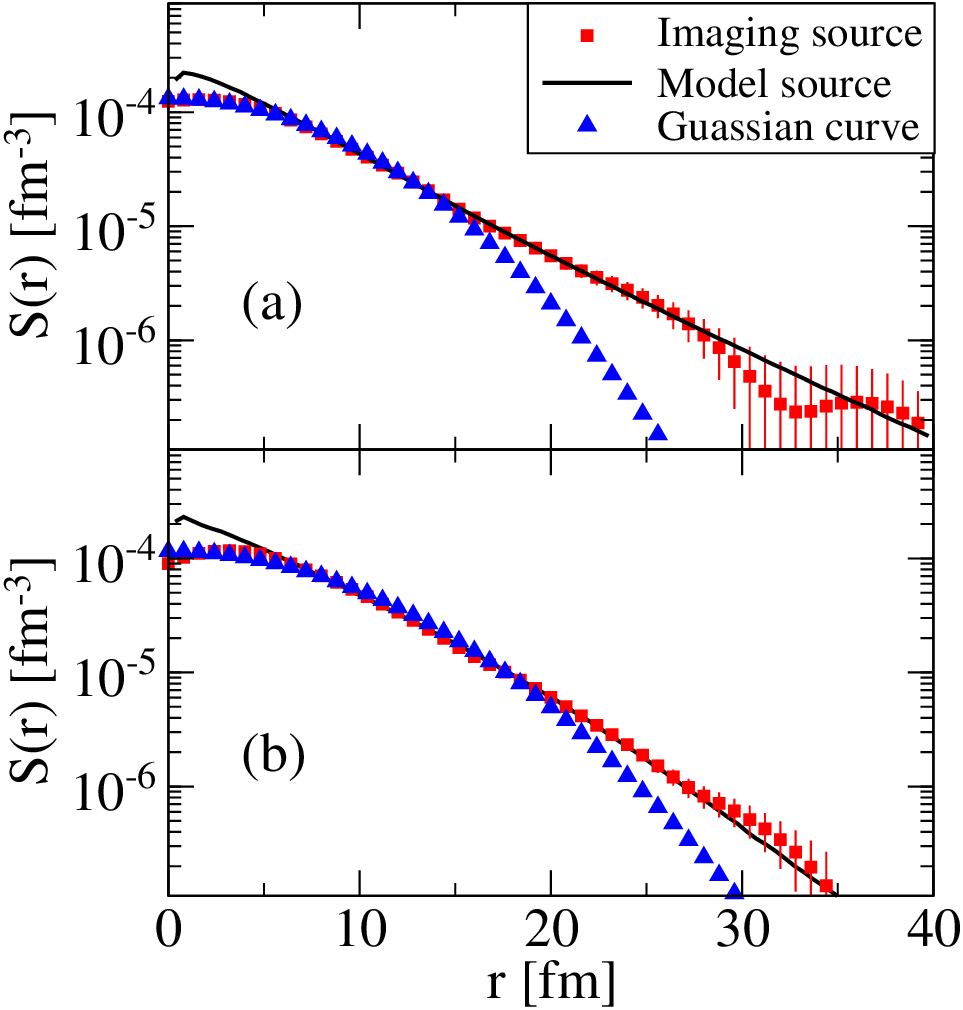}
\caption{(color online). Comparison of effects of long tail between the results in the CMFP(a) and CMFS(b) with the AMPT model. } \label{3}
\end{figure}

We can conclude that two factors may have led to the long tail appearing in Ref.\cite{PHE07}: one is a non-Gaussian pion-emitting source,and  the other is the imaging technique used in the CMFP. To show their respective influences, Fig.\ref{3} presents source functions obtained in various reference frames using the string melting multi-phase transport(AMPT) model \cite{ZWL05}. The results correspond to Au+Au collisions at $\sqrt{s_{_{NN}}}=200$ GeV with an impact parameter of $b=0$ fm
, using identical $\pi^{+}$ pairs to construct the correlation functions. Fig.\ref{3}(a) presents the results for the $200$ simulated events in the CMFP. The imaging source exhibits a more prominent tail than the Gaussian curve. This tail is caused by both the factors discussed above.  Fig.\ref{3}(b) presents the results for $10000$ simulated events in the CMFS(because we only choose pion pairs with $\bigtriangleup
E<10$~MeV, more statistics are needed). The imaging source under the CMFS exhibited a significantly smaller spatial tail than that under the CMFP reference frame. This tail is caused only by the non-Gaussian pion-emitting source.

\section{Summary}

In this work, we have identified and resolved a critical limitation in the pion-emitting source analysis related to reference frame selection. By systematically comparing the long-range source component in both the CMFP and CMFS frames, we discovered that the conventional imaging technique artificially generates an extended tail in the CMFP frame  even for a intrinsically Gaussian source. This artifact stems from kinematic distortions introduced by frame transformation. To address this fundamental issue, we have developed a novel imaging framework that operates directly in the CMFS frame, where pion pairs exhibit minimal kinematic correlations. Unlike previous energy-based pair selection methods, which implicitly assume a source morphology, our approach is model-independent and reconstructs the source function without ad hoc assumptions. This enables a more direct and unambiguous extraction of the underlying source structure. Validation against model-generated source functions confirms that our method substantially suppresses the non-Gaussian tail and reproduces the true source morphology with significantly higher fidelity. We thus propose that applying this refined imaging technique to existing and future experimental data will open new avenues for investigating source dynamics  potentially elucidating the physical mechanisms responsible for observed long-range correlations in heavy-ion collisions.

\end{document}